\documentclass[twocolumn,floatfix,times,trackchanges]{aastex63}

\def\sun{\hbox{$\odot$}}
\def\earth{\hbox{$\oplus$}}
\def\lesssim{\mathrel{\hbox{\rlap{\hbox{%
 \lower4pt\hbox{$\sim$}}}\hbox{$<$}}}}
\def\gtrsim{\mathrel{\hbox{\rlap{\hbox{%
 \lower4pt\hbox{$\sim$}}}\hbox{$>$}}}}

\def\arcmin{\hbox{$^\prime$}}
\def\arcs{\hbox{$^{\prime\prime}$}}

\def\farcs{\hbox{$.\!\!^{\prime\prime}$}}

\def\micron{\hbox{$\mu$m}}

\newcommand{\percbcm}{\mbox{~cm$^{-3}$}}
\newcommand{\persqcm}{\mbox{~cm$^{-2}$}}

\newcommand{\mJyperbeam}{\mbox{~mJy~beam$^{-1}$}}
\newcommand{\muJyperbeam}{\mbox{~$\mu$Jy~beam$^{-1}$}}
\newcommand{\kmpers}{\mbox{~km~s$^{-1}$}}
\newcommand{\vLSR}{$v_\mathrm{LSR}$}

\newcommand{\frest}{$f_\mathrm{rest}$}

\newcommand{\Eu}{$E_\mathrm{u}$}

\newcommand{\vwidth}{$\delta v$}

\newcommand{\Trot}{$T_\mathrm{rot}$}
\newcommand{\Ntot}{$N_\mathrm{tot}$}
\newcommand{\Lsun}{$L_\odot$}

\newcommand{\Lbol}{$L_\mathrm{bol}$}
\newcommand{\Tbol}{$T_\mathrm{bol}$}

\newcommand{\REVI}[1]{\textbf{#1}}
\newcommand{\REVII}[1]{#1}
\newcommand{\REVIII}[1]{#1}

\usepackage{CJKutf8}
\usepackage{graphicx}
\usepackage{longtable} \LTcapwidth=\textwidth
\usepackage{multirow}
\usepackage{subfigure}
\usepackage{ulem}
\usepackage{soul}
\usepackage{lipsum, babel}
\usepackage{enumitem}
\usepackage{amsmath}
\usepackage{euscript}
\usepackage{verbatim}
\usepackage{hyperref}

\setstcolor{red}

\submitjournal{ApJ}

\shorttitle{HOPS 87}
\shortauthors{Hsu et al.}
\graphicspath{{./}{figures/}}

\begin{document}

\begin{CJK*}{UTF8}{bsmi}
\title{ALMASOP. The Localized and Chemically rich Features near the Bases of the Protostellar Jet in HOPS 87}

\author[0000-0002-1369-1563]{Shih-Ying Hsu}
\email{seansyhsu@gmail.com}
\affiliation{Institute of Astronomy and Astrophysics, Academia Sinica, No.1, Sec. 4, Roosevelt Rd, Taipei 106216, Taiwan (R.O.C.)}

\author[0000-0002-3024-5864]{Chin-Fei Lee}
\email{cflee@asiaa.sinica.edu.tw}
\affiliation{Institute of Astronomy and Astrophysics, Academia Sinica, No.1, Sec. 4, Roosevelt Rd, Taipei 106216, Taiwan (R.O.C.)}

\author[0000-0012-3245-1234]{Sheng-Yuan Liu}
\affiliation{Institute of Astronomy and Astrophysics, Academia Sinica, No.1, Sec. 4, Roosevelt Rd, Taipei 106216, Taiwan (R.O.C.)}

\author[0000-0002-6773-459X]{Doug Johnstone}
\affiliation{NRC Herzberg Astronomy and Astrophysics, 5071 West Saanich Rd, Victoria, BC, V9E 2E7, Canada}
\affiliation{Department of Physics and Astronomy, University of Victoria, Victoria, BC, V8P 5C2, Canada}

\author[0000-0002-5286-2564]{Tie Liu}
\affiliation{Key Laboratory for Research in Galaxies and Cosmology, Shanghai Astronomical Observatory, Chinese Academy of Sciences, 80 Nandan Road, Shanghai 200030, People’s Republic of China}

\author[0000-0002-7287-4343]{Satoko Takahashi}
\affiliation{National Astronomical Observatory of Japan, National Institutes of Natural Sciences, 2-21-1 Osawa, Mitaka, Tokyo 181-8588, Japan}
\affiliation{Department of Astronomical Science, The Graduate University for Advanced Studies, SOKENDAI, 2-21-1 Osawa, Mitaka, Tokyo 181-8588, Japan}

\author[0000-0002-9574-8454]{Leonardo Bronfman}
\affiliation{Departamento de Astronom\'{i}a, Universidad de Chile, Casilla 36-D, Santiago, Chile}

\author[0000-0002-9774-1846]{Huei-Ru Vivien Chen}
\affiliation{Department of Physics and Institute of Astronomy, National Tsing Hua University, Hsinchu, 30013, Taiwan}

\author[0000-0002-2338-4583]{Somnath Dutta}
\affiliation{Institute of Astronomy and Astrophysics, Academia Sinica, No.1, Sec. 4, Roosevelt Rd, Taipei 106216, Taiwan (R.O.C.)}

\author[0000-0002-5881-3229]{David J. Eden}
\affiliation{Armagh Observatory and Planetarium, College Hill, Armagh, BT61 9DB, UK}

\author[0000-0001-5175-1777]{Neal J. Evans II}
\affiliation{Department of Astronomy, The University of Texas at Austin, 2515 Speedway, Stop C1400, Austin, Texas 78712-1205, USA}

\author[0000-0001-9304-7884]{Naomi Hirano}
\affiliation{Institute of Astronomy and Astrophysics, Academia Sinica, No.1, Sec. 4, Roosevelt Rd, Taipei 106216, Taiwan (R.O.C.)}

\author[0000-0002-5809-4834]{Mika Juvela}
\affiliation{Department of Physics, P.O.Box 64, FI-00014, University of Helsinki, Finland}

\author[0000-0002-4336-0730]{Yi-Jehng Kuan}
\affiliation{Department of Earth Sciences, National Taiwan Normal University, Taipei, Taiwan (R.O.C.)}
\affiliation{Institute of Astronomy and Astrophysics, Academia Sinica, No.1, Sec. 4, Roosevelt Rd, Taipei 106216, Taiwan (R.O.C.)}

\author[0000-0003-4022-4132]{Woojin Kwon}
\affiliation{Department of Earth Science Education, Seoul National University, 1 Gwanak-ro, Gwanak-gu, Seoul 08826, Republic of Korea}
\affiliation{SNU Astronomy Research Center, Seoul National University, 1 Gwanak-ro, Gwanak-gu, Seoul 08826, Republic of Korea}
	
\author[0000-0002-3179-6334]{Chang Won Lee}
\affiliation{Korea Astronomy and Space Science Institute (KASI), 776 Daedeokdae-ro, Yuseong-gu, Daejeon 34055, Republic of Korea}
\affiliation{University of Science and Technology, Korea (UST), 217 Gajeong-ro, Yuseong-gu, Daejeon 34113, Republic of Korea}

\author[0000-0003-3119-2087]{Jeong-Eun Lee}
\affiliation{Department of Physics and Astronomy, Seoul National University, 1 Gwanak-ro, Gwanak-gu, Seoul 08826, Korea}

\author[0000-0003-1275-5251]{Shanghuo Li}
\affiliation{Max Planck Institute for Astronomy, Königstuhl 17, D-69117 Heidelberg, Germany}

\author[0000-0002-1624-6545]{Chun-Fan Liu}
\affiliation{Institute of Astronomy and Astrophysics, Academia Sinica, No.1, Sec. 4, Roosevelt Rd, Taipei 106216, Taiwan (R.O.C.)}

\author[0000-0001-8315-4248]{Xunchuan Liu}
\affiliation{Shanghai Astronomical Observatory, Chinese Academy of Sciences, Shanghai 200030, PR China}

\author[0000-0003-4506-3171]{Qiuyi Luo}
\affiliation{Shanghai Astronomical Observatory, Chinese Academy  of Sciences, Shanghai 200030, People’s Republic of China}
\affiliation{School of Astronomy and Space Sciences, University of Chinese Academy of Sciences, No. 19A Yuquan Road, Beijing 100049, People’s Republic of China} 
\affiliation{Key Laboratory of Radio Astronomy and Technology, Chinese Academy of Sciences, A20 Datun Road, Chaoyang District, Beijing, 100101, P. R. China}

\author[0000-0003-2302-0613]{Sheng-Li Qin}
\affiliation{Department of Astronomy, Yunnan University, and Key Laboratory of Astroparticle Physics of Yunnan Province, Kunming, 650091, People's Republic of China}

\author[0000-0002-4393-3463]{Dipen Sahu}
\affiliation{Physical Research laboratory, Navrangpura, Ahmedabad, Gujarat 380009, India}
\affiliation{Academia Sinica Institute of Astronomy and Astrophysics, 11F of AS/NTU Astronomy-Mathematics Building, No.1, Sec. 4, Roosevelt Rd, Taipei 106216, Taiwan, R.O.C.}

\author[0000-0002-7125-7685]{Patricio Sanhueza}
\affiliation{National Astronomical Observatory of Japan, National Institutes of Natural Sciences, 2-21-1 Osawa, Mitaka, Tokyo 181-8588, Japan}
\affiliation{Department of Astronomical Science, The Graduate University for Advanced Studies, SOKENDAI, 2-21-1 Osawa, Mitaka, Tokyo 181-8588, Japan}

\author[0000-0001-8385-9838]{Hsien Shang (尚賢)}
\affiliation{Institute of Astronomy and Astrophysics, Academia Sinica,  Taipei 106216, Taiwan}

\author[0000-0002-8149-8546]{Kenichi Tatematsu}
\affiliation{Nobeyama Radio Observatory, National Astronomical Observatory of Japan, National Institutes of Natural Sciences, 462-2 Nobeyama, Minamimaki, Minamisaku, Nagano 384-1305, Japan}
\affiliation{Department of Astronomical Science, The Graduate University for Advanced Studies, SOKENDAI, 2-21-1 Osawa, Mitaka, Tokyo 181-8588, Japan}

\author[0000-0001-8227-2816]{Yao-Lun Yang}
\affiliation{Star and Planet Formation Laboratory, RIKEN Cluster for Pioneering Research, Wako, Saitama 351-0198, Japan}


\begin{abstract} 
HOPS 87 is a Class 0 protostellar core known to harbor an extremely young bipolar outflow and a hot corino. We report the discovery of localized, chemically rich regions near the bases of the two-lobe bipolar molecular outflow in HOPS 87 containing molecules such as H$_2$CO, $^{13}$CS, H$_2$S, OCS, and CH$_3$OH, the simplest complex organic molecule (COM).
The locations and kinematics suggest that these localized features are due to jet-driven shocks rather than being part of the hot corino region encasing the protostar.
The COM compositions of the molecular gas in these jet-localized regions are relatively simpler than those in the hot corino zone. 
We speculate that this simplicity is due to either the liberation of ice with a less complex chemical history or the effects of shock chemistry. 
Our study highlights the dynamic interplay between the protostellar bipolar outflow, disk, inner core environment, and the surrounding medium, contributing to our understanding of molecular complexity in solar-like young stellar objects.
\end{abstract}


\section{Introduction}
\label{sec:Intro}
Interstellar complex organic molecules (COMs or iCOMs) found within low-mass young stellar objects (YSOs) are significant in astronomy due to their potential role in prebiotic chemistry in protoplanetary systems. 
The presence of warm gaseous COMs in YSOs is often explained by the ``thermal desorption'' paradigm \citep[e.g., ][]{2007Garrod_reactive_desorption,2009Herbst_COM_review}. 
Initially, during the starless core phase, with a temperature of $\sim$10 K, CO molecules are primarily depleted by freeze-out and trapped in the icy mantles coating dust grains. 
Through grain-surface reactions, such as hydrogenation, COMs gradually form within these CO-rich ices. 
As the core evolves into the protostellar stage, the central protostar heats up the surrounding envelope, causing the desorption of these previously trapped COMs. 
In the innermost envelope, where temperatures exceed the ice sublimation threshold of $\sim$100 K, the majority of the trapped COMs are released into the gas of the newly formed hot corino \citep{2004Ceccarelli_HotCorino}. 
Hot corinos are commonly observed in protostellar cores, provided that their innermost envelopes attain temperatures of at least 100 K and guarantee ice sublimation \citep{2004Ceccarelli_HotCorino, 2023Hsu_ALMASOP}.
Furthermore, protostellar outbursts may extend the boundaries of thermal desorption due to the rapid increase of the temperature in the disk \citep{2019Lee_V883-Ori_outburst}. 

In addition to the heating from the central protostar, shocks also play a role in releasing COMs in the gas phase. 
For example, COM factories have been discovered in the protostellar outflow L1157, particularly in the B1 shock region \citep[e.g., ][]{2008Arce_L1157-B1_COMs,2014Mendoza_L1157-B1_COMs,2015Codella_L1157-B1_COMs,2017Lefloch_L1157-B1_COMs}. 
\citet{2008Arce_L1157-B1_COMs} suggested that COMs are primarily formed on the grain surfaces and then expelled from icy mantles by shocks. 
For [BHB2007] 11, a protobinary protostellar system, \citet{2022Vastel_BHB2007-11_COMs} suggested that the presence of hot methanol is due to shocks possibly generated by streamers towards the cores, impacting the quiescent gas within the circumbinary envelope, circumbinary disk, or the two circumstellar disks.
Similarly, in the nearly edge-on protostellar disk around HH-212 \citep[e.g., ][]{2016Codella_HH212_H2O_COM,2017Lee_HH212,2019Lee_HH212_COM_atm}, some COMs near the disk edge may be released by the heat generated by ``accretion shocks" \citep{2017Lee_HH212,2021Tabone_HH212_cavity}, while others away from the disk edge could be released by the heat obtained from the radiation of the central protostar and/or disk-wind interaction \citep{2022Lee_HH212_stratification}. 
A similar accretion shock mechanism was proposed by \citet{2020Oya_IRAS16293-2422-A_few_au} based on the steep rise in H$_2$CS rotation temperature to 300~K or higher at a radius of 50~au from the protostar in IRAS 16293–2422 A. 
This localized heating mechanism received further support from the detection of a elevated ring-like warm structure in the Class 0 protostellar core B335 \citep{2022Okoda_B335_few_au}.

These findings underscore the potential and significance of shocks in investigating COMs in YSOs, as they are potentially capable of releasing COMs from the ice phase to the gas phase. 
Furthermore, chemically rich shocked regions may exhibit distinct chemical characteristics compared to hot corinos where the ice is more gradually sublimated. 
For instance, certain COMs such as CH$_3$CHO are observed to be more prevalent in the L1175-B1 shocked region than in hot corinos, with abundance ratios ranging from 2 to 10, depending on the species \citep{2017Lefloch_L1157-B1_COMs}. 
Additionally, some studies suggest that gas-phase reactions are enhanced following shocks \citep{2015Codella_L1157-B1_COMs}. 
The limited identification of chemically rich shocked regions leads to a desire for further discoveries and in-depth investigations, particularly targeting those YSOs hosting abundant COMs. 

HOPS 87 \citep{2016Furlan_HOPS}, also known as OMC-3 MMS 6 \citep{1997Chini_OMC23_dust} and G208.68-19.20N1 \citep{2020Dutta_ALMASOP}, is a Class~0 protostellar core in the Orion Molecular Cloud 3 (OMC-3) region with a bolometric luminosity \Lbol\ of $38\pm13$~\Lsun, a bolometric temperature \Tbol\ of $36.7\pm14.5$~K, and local standard of rest velocity \vLSR\ of $11.4$~\kmpers\ \citep{2020Dutta_ALMASOP}.
This source is one of the 11 protostellar cores with a detected hot corino reported by the ``ALMA Survey of Orion PGCCs (ALMASOP)'' project \citep{2020Hsu_ALMASOP,2022Hsu_ALMASOP}, where the ALMA is Atacama Large Millimeter/submillimeter Array and the PGCCs are Planck Galactic Cold Clumps \citep{2016Planck_PGCC}. 
It is the third-most luminous protostellar core in the catalogued sample. 
There are several COM species detected toward this source, including CH$_3$OH, CH$_2$DOH, $^{13}$CH$_3$OH, CH$_3$CHO, HCOOCH$_3$, and NH$_2$CHO. 

\citet{2012Takahashi_HOPS87_outflow} reported an extremely young outflow associated with HOPS 87 (MMS6) with an estimated outflow dynamic range of $\le$100~yr. 
The authors reported an inclination of 45\degr\ for the bipolar outflow by comparing their position-velocity (PV) diagram with a kinetic outflow model \citep{2000Lee_jet_wind_model}. 
\citet{2019Takahashi_HOPS87_outflow} later estimated the inclination angle to be 60\degr\ (roughly egde-on) based on comparisons between the observed toroidal magnetic field and numerical calculations. 
\citet{2012Takahashi_HOPS87_outflow} reported a detection of HCN ($J=4-3$), tracing an increase in line width at the end of the molecular outflow, which indicates a clear bow-shock type velocity structure. 
The jets in HOPS 87 were recently studied by \citet{2024Dutta_ALMASOP_jet}, and the projected velocity of the jets was reported to be $43\pm6$~\kmpers. 
Assuming that the majority of low-velocity components in the bipolar outflow are wind components, the authors estimated an inclination angle of the bipolar outflow to be 65\degr. 

In this paper, we report newly discovered, localized, chemically rich regions associated with jet-driven shocks near the two bases of the bipolar outflow in the protostellar core HOPS 87.
In \S \ref{sec:Obs}, we introduce the two observation programs used for this study. 
In \S \ref{sec:mom0}, we show the (integrated) intensity maps of selected molecular transitions, which reveal the two-lobe, localized, and chemically rich regions near the bases of the bipolar molecular outflow. 
Moreover, in \S \ref{sec:kinematics}, we investigate the kinematics in the localized regions. 
In \S \ref{sec:Disc} and \S \ref{sec:Conclusions}, we discuss our findings and make summarize our conclusions, respectively.


\begin{figure*}[htb!]
\centering
\includegraphics[width=0.96\textwidth]{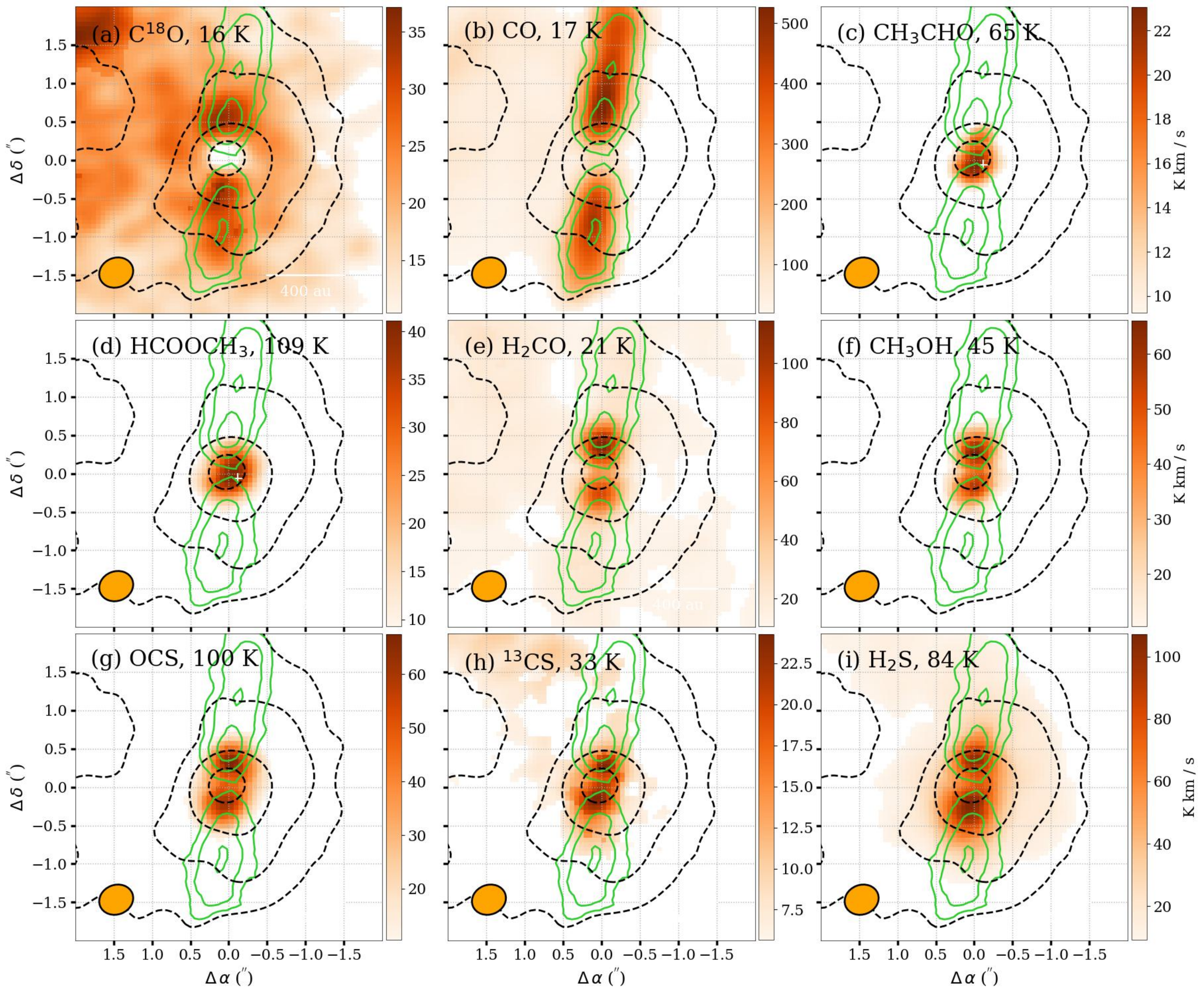}
\caption{\label{fig:mom0_outflow} 
Integrated intensity images (moment-0 maps) of selected molecular transitions: C$^{18}$O, CO, CH$_3$CHO, HCOOCH$_{3}$, H$_{2}$CO, CH$_{3}$OH, OCS, $^{13}$CS, and H$_2$S.  
The origin of the coordinate system is the 1.3~mm continuum peak at [$\alpha_{2000}, \delta_{2000}$] $=$ [05$^\mathrm{h}$35$^\mathrm{m}$23$^\mathrm{s}$\!\!.42, -05$\degr$01$\arcmin$30$\arcs$\!\!.6], and the ranges of right ascension (x-axis) and declination (y-axis) span $\pm2\arcs$.
Each panel is labeled at the top with the corresponding chemical formula and the upper energy level of the transition shown in color.
The velocity ranges for integration are $\pm 40$~\kmpers\ for CO and SiO, $\pm 3$~\kmpers\ for $^{13}$CS, and $\pm 7.5$~\kmpers for others. 
The black dashed contours depict the 1.3~mm continuum, set at levels of [5, 10, 20, 40, 80, 160]$\sigma$, where $\sigma$ corresponds to 0.9\mJyperbeam\ (or 0.13~K in terms of brightness temperature).
The orange ellipse located at the bottom left corner indicate the beam sizes for the Cycle 6 ALMA program.
The green contours display the SiO integrated intensity map, highlighted at levels of [10, 25, 40]$\sigma$, where each $\sigma$ corresponds to 28\mJyperbeam\kmpers. 
The distribution of the molecules in panels (e)--(i) are particularly noteworthy; these molecules are not typical outflow tracers but exhibit two-lobed distributions.
The white cross in panels (c) and (d) illustrate the peak position of the corresponding transition. 
}
\end{figure*}

\section{Observations}
This study mainly retrieves data obtained from the ALMASOP program (\#2018.1.00302.S) and supplemental data obtained from an earlier archival ALMA project (\#2015.1.00341.S).

\label{sec:Obs}
\subsection{ALMASOP Project: ALMA Cycle 6 \#2018.1.00302.S}
The ALMASOP data were obtained from ALMA Cycle~6 program \#2018.1.00302.S (PI: Tie Liu) in Band~6 (1.3~mm or 230~GHz) with the combination of both the 12-m array (configurations C43–5 and C43–2) and 7-m array (also known as Atacama Compact Array, ACA, or Morita Array). 
The projected baselines ranged from 4.7 to 1066 k$\lambda$, and the resulting maximum recoverable scale was approximately 26\arcs.
This survey selected 72 clumps in the Orion A, B, and $\lambda$ Orionis clouds as targets, beginning from a sample of PGCCs \citep{2016Planck_PGCC}. 
The four spectral windows centered at 216.6, 218.9, 231.0, and 233.0 GHz have a uniform bandwidth of 1,875~MHz with a spectral resolution of 1.129~MHz (1.4~\kmpers\ at 230~GHz). 

We used \texttt{tclean} in CASA \citep[][]{casa:2022} with a robust value of 2.0. 
The resulting angular resolution was $\sim$ 0\farcs{42} (which corresponds to about 168 au for a distance of $\sim400$~pc) for both continuum and continuum-subtracted data cubes, and the sensitivities of the image reach $\sim$12~\muJyperbeam\ and $\sim$3.3~\mJyperbeam\ for the full--band continuum and each channel, respectively. 
For more details about the ALMASOP project, please refer to \citet{2020Dutta_ALMASOP}. 

\subsection{Archival Data: ALMA Cycle 3 \#2015.1.00341.S}

This study also retrieved ALMA Band 6 data obtained from a Cycle~3 program (\#2015.1.00341.S, PI: Satoko Takahashi) for better-visualizing the intensity maps of the CO $2-1$ and CH$_3$OH $5_{1,4}$ -- $4_{2, 4}$ transitions as it had a much better spatial resolution (FWHM beam size $\sim$ 0\farcs{16}).  
These observations were conducted on 2016 September 18. 
The data cubes used in this study have a spectral resolution of 0.244~MHz, corresponding to a velocity resolution of 0.32~\kmpers. 
The projected baseline ranged from 16 to 3200 k$\lambda$, and the maximum recoverable scale was about 1\arcs. 

The images exported by the Additional Representative Images for Legacy (ARI-L) project \citep{2021Massardi_ARI-L} were utilized. 
ARI-L, endorsed by the Joint ALMA Observatory (JAO) and the European Southern Observatory (ESO), is a European Development project for ALMA. 
Its primary objective is to employ the ALMA Imaging Pipeline on data from initial observing cycles (2–4) to regenerate data products with equivalent completeness and quality to those currently produced by ALMA for recent observations. 
The original CASA pipeline version was 4.7.0 and the Briggs robust parameter was set to 0.5.
With a FWHM beam size of 0\farcs{16}, the sensitivity achieved $\sim$6~\mJyperbeam\ for a channel width of 0.244~MHz.

\begin{figure*}[htb!]
\centering
\includegraphics[width=0.99\textwidth]{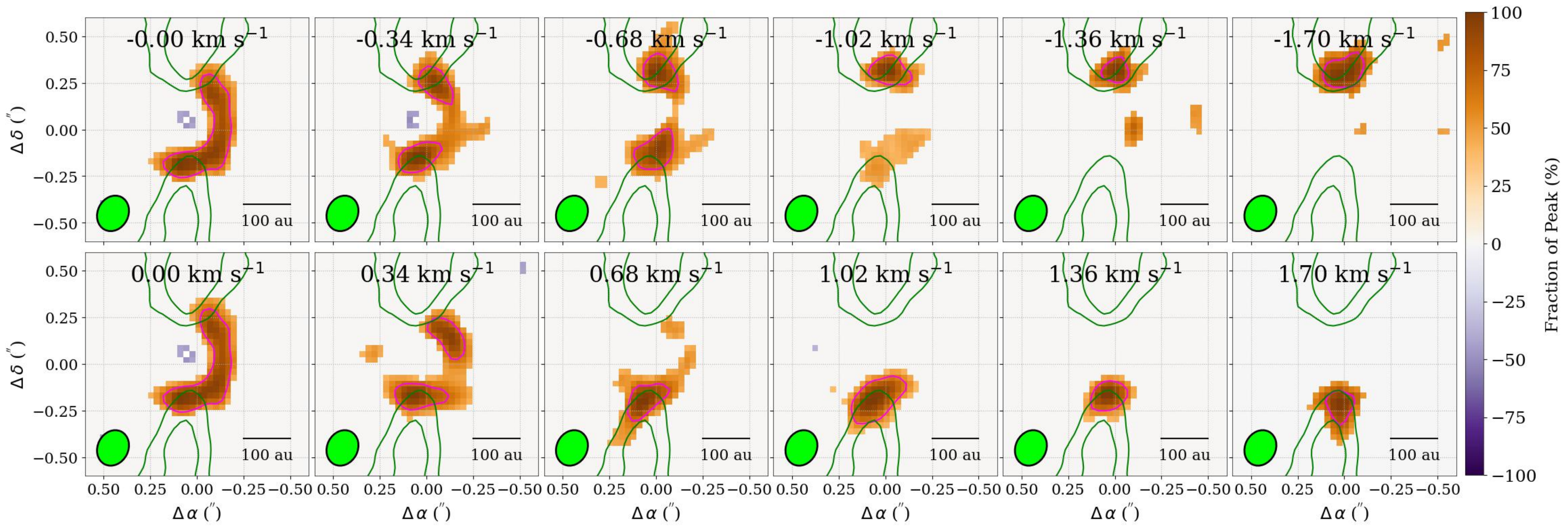}
\caption{\label{fig:chn_CH3OH_Cycle3} 
The CH$_3$OH $5_{1,4}-4_{2,3}$ transition in Cycle 3 observations. 
The top and bottom rows show the images at the blue- and red-shifted channels, respectively. 
The labels in each panel shows the relative velocity ($\Delta v = v - v_\mathrm{LSR}$). 
At $\Delta v=0$~\kmpers\ the intensity map displays an arc-like emission pattern west of the center with an absorption feature toward the continuum peak. 
At $|\Delta v|>0.68$~\kmpers, there is a single lobe ($\Delta \delta \sim 0\farcs{3}$) on each of the northern and the southern sides of the arc ($\Delta \delta \sim 0\farcs{2}$) seen at $\Delta v=0$~\kmpers. 
}
\end{figure*}


\section{Results}

\subsection{Dust Continuum and Molecular Line Emission}
\label{sec:mom0}

Table~\ref{tab:molec_info} \REVII{lists} the information of the molecular transitions used in this study.  
Figure~\ref{fig:mom0_outflow} shows the integrated intensity images of selected molecular transitions within a size of 4\farcs{0} ($\sim$1600~au) overlaid by 1.3~mm continuum and the SiO integrated intensity maps. 
For completeness, the integrated intensity images of all the molecular transitions in this study \REVII{are shown in Figure~\ref{fig:mom0_all}. }

\subsubsection{1.3 mm Dust Continuum}
\label{sec:mom0:cont}

Figure~\ref{fig:mom0_outflow} shows the observed 1.3~mm dust continuum which appears to trace the extended envelope. 
As illustrated by the dashed black contours, the continuum is spherically-symmetric at the small scale ($\sim$150~au), becomes elongated at a middle scale ($\sim$400~au), and extend to north and east at a large scale ($\sim$600~au). 
The brightness temperature of the 1.3 mm continuum peak is 47 K with a beam size of 0\farcs{42}. 
\citet{2012Takahashi_HOPS87_env} and \citet{2019Takahashi_HOPS87_outflow} reported brightness temperatures of 52 K and 192 K with beam sizes of 0\farcs{27} and 0\farcs{024}, respectively. 
These relatively high \REVII{dust continuum brightness temperatures} suggest that the inner region is very warm, similar with the cases of NGC 1333 IRAS 4A1 (57~K) and 4A2 (42~K) reported by \citet{2019Sahu_IRAS4A_hot-corino-atm}. 
In addition, HOPS 87 and IRAS 4A2 are reported to harbor \REVII{hot corinos} \citep[e.g., ][]{2022Hsu_ALMASOP,2019Sahu_IRAS4A_hot-corino-atm}. 
Due to these similarities, we adopt the assumed dust temperature (60~K) from \citet{2019Sahu_IRAS4A_hot-corino-atm} for evaluating the optical depth and the molecular hydrogen column density. 
We note that the beam size of our study ($\sim 160$ au) is larger than the value ($\sim 80$ au) of \citet{2019Sahu_IRAS4A_hot-corino-atm}. 
As a result, this adopted continuum brightness temperature (60~K) could be overestiamted. 
In such case, the derived optical depth and the molecular hydrogen column density in the following contexts are their lower limits. 

We evaluate the optical depth ($\tau_\nu$) using the following equation adopted from \citet{2008Kauffmann_MAMBO}: 
\begin{equation}
    I_\nu = B_\nu(T_\mathrm{d})(1-e^{-\tau_\nu}) = \frac{F_\nu^\mathrm{beam}}{\Omega_\mathrm{A}}, 
\end{equation}
where $I_\nu$ is the beam-averaged intensity, 
$B_\nu$ is the Planck function, 
$T_\mathrm{d}$ is the dust temperature, 
$F_\nu^\mathrm{beam}$ is the observed flux per beam, and 
$\Omega_\mathrm{A}$ is the solid angle of the beam. 
Adopting the brightness temperature at the continuum peak, the derived optical depth $\tau_\nu$ is 1.55, indicating that the continuum is somewhat optically thick. 

We estimate the molecular hydrogen column density $N(\mathrm{H}_2)$ adopting Equation A.9 in \citet{2008Kauffmann_MAMBO}: 
\begin{equation}
    N(\mathrm{H}_2) = \frac{\tau_\nu}{\mu\,m_\mathrm{H}\,\kappa_\nu}, 
\end{equation}
where $\mu$ is the molecular weight per hydrogen molecule $\sim$2.8, 
$m_\mathrm{H}$ is the mass of atomic hydrogen, 
and $\kappa_\nu$ is the dust mass opacity. 
To estimate $\kappa_\nu$, we adopt the form of $\kappa_\nu=0.1(\nu/\mathrm{THz})^\beta\,\mathrm{cm^2\,g^{-1}}$, where $\beta$ is the the dust opacity index, from  \citet{1990Beckwith_DustOpacity}. 
We assume the index $\beta$ to be 1.70, which is the typical opacity index of cold clumps in the submillimeter band \citep{2018Juvela_dustIndex}, and the estimated $\kappa_\nu$ is 0.0083 cm$^2$ g$^{-1}$ at 1.3~mm.
Assuming a gas-to-dust ratio of 100, the resulting molecular hydrogen column density $N(\mathrm{H}_2)$ is $2.89\times10^{25}$\persqcm. 

\begin{figure*}[htb!]
\centering
\includegraphics[width=0.96\textwidth]{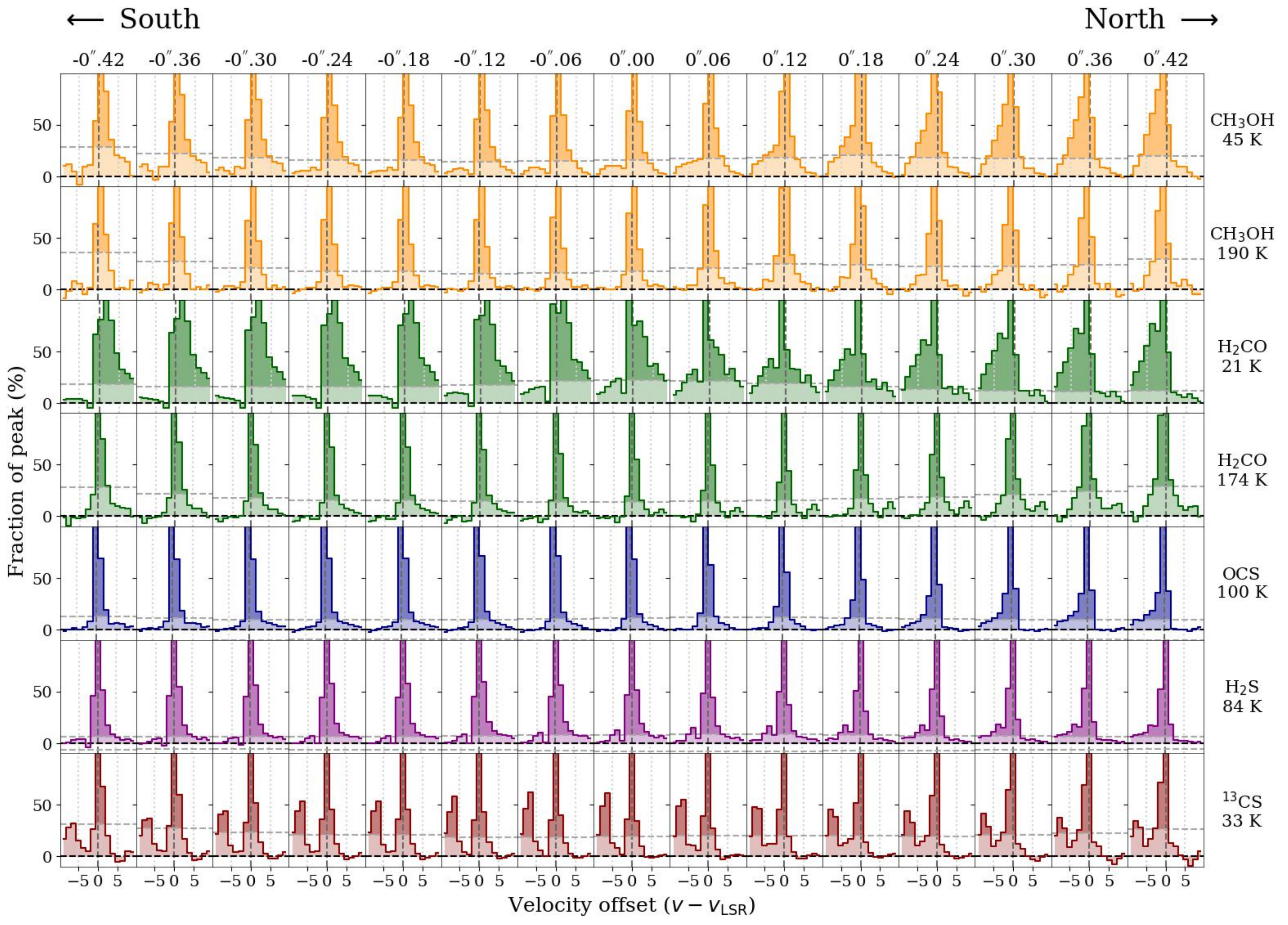}
\caption{\label{fig:spec_along_outflow_selected} Spectra of selected molecular transitions along the cavity axis. 
The labels at the right of each row show the formula and the upper energy of each transition. 
The labels at the top of each column show the angular offset from the continuum peak. 
The horizontal dashed lines represent the level of 5$\sigma$. 
In the panels of $^{13}$CS, the lines at $\sim -7$~\kmpers\ is another molecular transition. 
The y-axis is the fraction of peak intensity. 
The north is blue-shifted (coming toward us) and the south is red-shifted (moving away), and the lines extend to bluer and redder velocities at the north and south positions, respectively.
}
\end{figure*}

\subsubsection{C$^{18}$O: Extended Warm Gas}
C$^{18}$O is a good tracer of warm gas, as it is released into gas phase at a temperature of $\sim$20~K \citep{2021Tychoniec_chemical_tracer} and has a low critical density \citep[$\sim$10$^4$\percbcm\ for $J=2-1$, ][]{2001Saito_Centaurus_C18O}. 
We show in Figure~\ref{fig:mom0_outflow} (a) the C$^{18}$O integrated intensity map, which \REVII{overall extends} to the north and \REVII{east}, is consistent with the dust continuum. 
In addition to the extended gas, there appears to be a hole absent of emission centered at the continuum peak position with a size of 0\farcs{5}, which is comparable to the beam size. 
Given that the C$^{18}$O emission is significant around the hole, it is unlikely that it \REVII{results} from a complete absence of C$^{18}$O molecules. 
This hole could be due to the comparable dust (continuum) temperature and the C$^{18}$O gas (excitation) temperature.

At the north and south sides of the hole, the bright C$^{18}$O emission may result from the bipolar outflow, as the positions are corresponding to the outflow lobes illustrated by the green contours. 
Near the northeast corner ($\Delta\alpha\sim+1\farcs{5}$, $\Delta\delta\sim+1\farcs{5}$) of panel (a), a separated peak of C$^{18}$O integrated intensity is detected \REVII{and corresponds} to a separate source (MMS6-NE at 0.9~mm and IRS 3 at near-IR) identified as a Class~I protostellar core by \citet{2009Takahashi_MMS6}. 

\subsubsection{CO and SiO: Outflows}
We show in Figure~\ref{fig:mom0_outflow} (b) the integrated intensity map of the typical outflow tracer CO ($J=2-1$). 
Consistent with the discoveries of \citet{2012Takahashi_HOPS87_outflow} in CO ($3-2$), the bipolar outflow, which is compact (about 2\arcs\ long for each lobe), exhibits two lobes at the northern and southern sides. 
As we will show in the following sections, the blue-shifted and red-shifted characteristics are in the northern and southern lobes, respectively.
The absence of CO emission at the region below a $\Delta p$ (the position offset from the continuum peak position) of 0\farcs{2} is due to both absorption along the line of sight and missing flux at low velocities due to interferometry. 
For the former, the absorption feature is also confirmed in the line profile at the continuum peak position.  
The latter was anticipated as implied from the PV diagram shown in Sect. \ref{sec:kinematics}.

SiO is another well-known outflow gas tracer, particularly the protostellar jet and the consequent shocked and collision regions. 
As illustrated by the green contours \REVII{in Figure~\ref{fig:mom0_outflow}}, SiO also trace outflowing gas \REVII{in HOPS 87.} 
\REVII{The sizes of the northern and southern SiO lobes are comparable to those of the CO outflows. }

\subsubsection{CH$_3$CHO and HCOOCH$_3$: Hot Corino Zone}
\label{sec:mom0:hotcorino}

A hot corino is a warm region, particularly the innermost envelope \citep{2023Hsu_ALMASOP} \REVII{around a low-mass protostar}, abundant in saturated COMs such as CH$_3$CHO and HCOOCH$_3$. 
Figure~\ref{fig:mom0_outflow} (c) and (d) show the CH$_3$CHO 65~K and HCOOCH$_3$ 109~K integrated \REVII{intensity} images, respectively, which illustrate the compact hot corino zone previously reported by \citet{2020Hsu_ALMASOP} and \citet{2022Hsu_ALMASOP}. 
This hot corino zone resides in the hole seen in the C$^{18}$O map, supporting that the hole seen in the C$^{18}$O integrated intensity map likely does not result from \REVII{a} lack of C$^{18}$O gas. 

\REVII{As illustrated by the white crosses} in Figure~\ref{fig:mom0_outflow} (c) and (d), the bright COM emission is slightly offset toward the west from the continuum peak position. 
Figure \ref{fig:chn_CH3OH_Cycle3} shows the intensity maps at each channel of a CH$_3$OH transition obtained from the Cycle 3 observations
At $\Delta v=0$~\kmpers\ the intensity map displays an arc-like emission pattern west of the center with an absorption feature toward the continuum peak. 
This arc outlines where the hot corino is, which is the warm COM emission resided in the innermost envelope, where the C$^{18}$O emission shows the absorption feature. 
The existing COM emission also suggest that the temperatures of the COMs are higher than that of C$^{18}$O. 
In other words, the warm gas-phase COM molecules are distributed in both the central absorption region and the arc emission region, and the former is due to the even hotter continuum at the center.
As a result, the offset of the COM emission observed in our Cycle 6 program \REVII{results} from the asymmetric innermost warm envelope gas surrounding the central warmer continuum optically thick region. 
It is noteworthy that the distribution is different at higher velocities, which will be elaborated on in the following sections.
The COM absorption feature \REVII{we observed in CH$_3$OH} is also been seen in IRAS 4A1  \citep{2019Sahu_IRAS4A_hot-corino-atm} and IRAS 16293-2422 B \citep[e.g., ][]{2018Oya_IRAS16293B_subarcsec}. 

\subsubsection{Molecules Exhibiting Two Localized Lobes}

In Figure~\ref{fig:mom0_outflow}, some molecules, \REVII{namely H$_{2}$CO (panel e), OCS (panel g), $^{13}$CS (panel h), and H$_2$S (i), apparently exhibit two-lobe signatures}. 
Interestingly, CH$_{3}$OH \REVII{(panel f)} exhibits a two-lobe signature in addition to its hot corino localization at low velocities. 
These molecules are not ``typical'' outflow tracers, while H$_2$CO could also trace the low-velocity outflows \citep{2024Izumi_ASHES_HotGas}. 
The position offset from the continuum peak ($\Delta p$) of the lobes for different molecules are similar ($\sim$0\farcs{3}), indicating \REVII{that the two regions are chemically rich}. 
The sizes of the lobes are comparable to the size of the beam, indicating that these chemically rich regions are localized. 

These two chemically rich localized regions are near the bases of the bipolar SiO molecular outflow (green contours).
We first checked whether the two regions are the outer extended parts of the hot corino in HOPS 87. 
As shown in Figure~\ref{fig:chn_CH3OH_Cycle3}, the intensity maps at $\Delta v=0$~\kmpers\ and at $|\Delta v|>0.68$~\kmpers\ \REVII{display} different patterns. 
The former traces the hot corino zone, as we discuss in Sect. \ref{sec:mom0:hotcorino}, whereas the latter exhibts a single lobe on each of the northern and the southern sides of the hot corino zone. 
Particularly at $|\Delta v|<0.68$~\kmpers, the position offset of the lobe ($\Delta \delta\sim$0\farcs{3}) appears to be slightly \REVII{beyond} the ``boundary'' of the hot corino ($\Delta \delta\sim$0\farcs{2}) seen at $\Delta v=0$~\kmpers. 
These observations suggest that the two lobes and the hot corino are tracing different regions. 
Moreover, the positions of the lobes, along with their consistent velocity directions with the bipolar outflow imply the potential association of the molecular gas with the bipolar outflow. 
This could be similar to OCS in the high-mass protostellar core NGC 2264 CMM3B, which also exhibits two lobes at the bases of the outflows, as found by \citet{2024Shibayama_NGC2264CMM3}. The authors suggest that the OCS is tracing the bases of outflows in addition to the disk/envelope system.

\subsection{Kinematics}
\label{sec:kinematics}

To further justify the association between the bipolar outflow and the two localized chemically rich regions, we initially extract the continuum-subtracted spectra along the cavity axis (position angle PA $=-4$\degr), centered at the 1.3~mm continuum peak. 
Figure \ref{fig:spec_along_outflow_selected} depicts the spectra of selected molecular transitions along the outflow axis. 
Note that the y-axis is the fraction of peak intensity for better visualizing the line profiles of all pixels. 
Despite the limited spectral resolution ($\sim$1.3\kmpers), discernible symmetry between spectra at the north and south positions is observed. 
Firstly, the north is blue-shifted, coming toward us, and the south is red-shifted, moving away. 
Secondly, the lines extend to bluer and redder velocities at the north and south positions, respectively.
These consistencies solidify their associations with the bipolar outflow. 
The velocity of the CO outflow reaches 40~\kmpers \citep{2020Dutta_ALMASOP}, while the velocities of \REVII{the molecules in Figure~\ref{fig:spec_along_outflow_selected}} are around 5~\kmpers. 

\begin{figure*}[htb!]
\centering
\includegraphics[width=.99\textwidth]{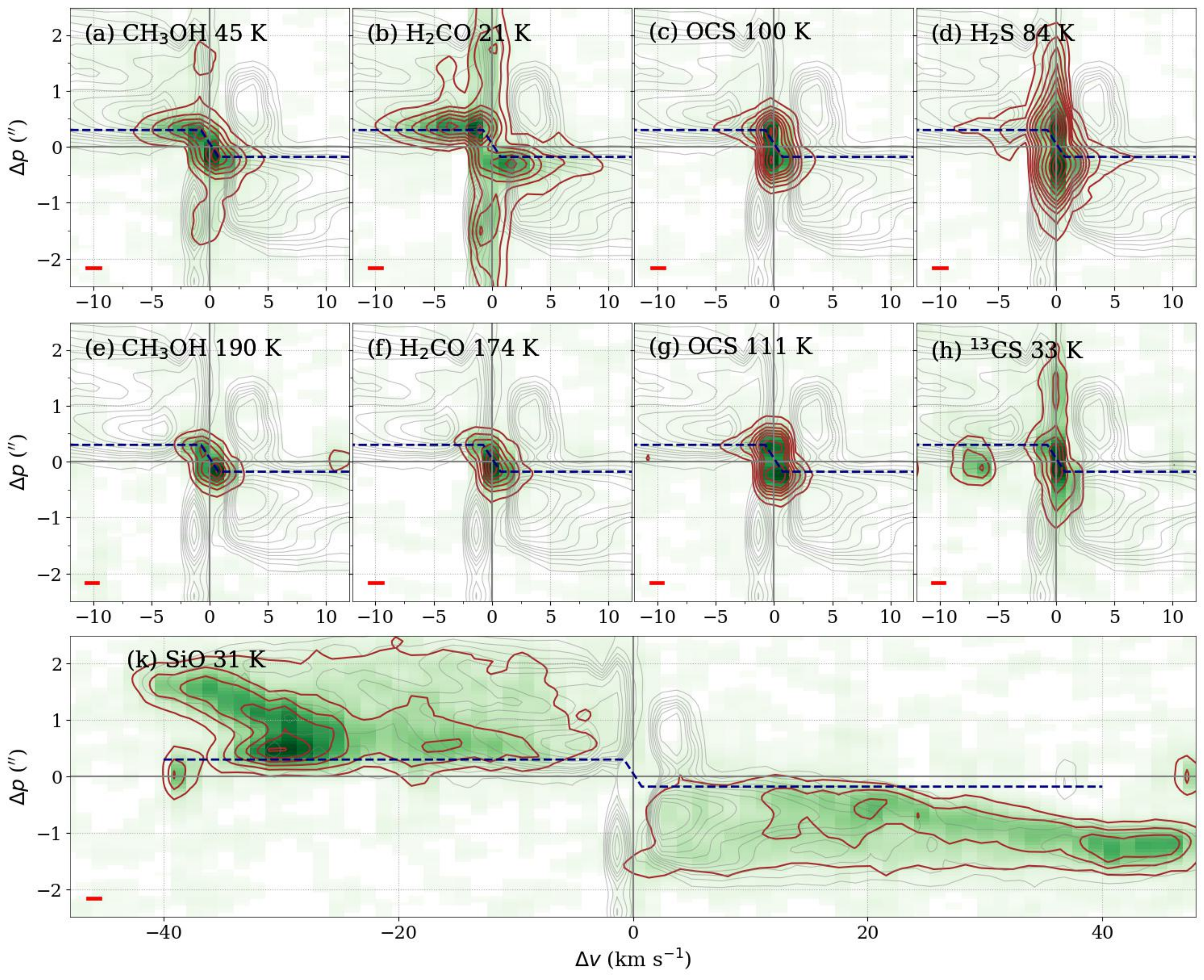}
\caption{\label{fig:PVtranss} 
The PV diagrams of selected molecular transitions along \REVII{the} outflow axis (position angle PA~$=-4$\degr \REVII{with} the center is at the 1.3~mm continuum peak. 
The relative position \REVII{offset} ($\Delta p$) is from south ($\Delta p<0\arcs$) to north ($\Delta p>0\arcs$). 
The upper left label in each panel show the chemical species and the upper energy \Eu\ of the transitions. 
Note that the diagrams are exported from the data cube of the full spectral window so the \REVII{separated components} may \REVII{come} from molecular transitions other than the labeled one in each label (e.g., the components at (-7.5\kmpers, 0\arcs) in panel (h) and at (-7.5\kmpers, 0\arcs) in panel (k).
The color maps and the brown contours are the PV diagram of the labeled transitions.
The grey contours represent the CO PV diagram. 
The brown and grey contours are in steps of [5, 10, 15, 20, 25, 30, 40, 50, 60, 70, 80]$\sigma$. 
The navy polyline, referred from the PV diagram of the CH$_3$OH~45~K transition, consists of two horizontal lines at $\Delta p=0\farcs{3}$ and $-0\farcs{2}$ and a \REVII{diagonal} line moves  from $\Delta p=0\farcs{3}$ to $-0\farcs{2}$ at $\Delta v=0.75$\kmpers\ and $-0.75$\kmpers, respectively. 
The diagrams are exported by CARTA 4.0 \citep[][]{2021Comrie_CARTA}. 
The gap in CO indicates the missing flux at low velocities due to interferometry. 
\REVII{
The broadening of the velocity range localized at $\Delta p=0\farcs{3}$ and $-0\farcs{2}$ is significant in CH$_3$OH, H$_2$CO, OCS, and H$_2$S and tentatively observed in $^{13}$CS. 
Additionally, the rapid growth of velocity $\Delta v$ with position offset $\Delta p$ described in \citet{2001Lee_jet_wind_model} is significant in SiO. 
}
These characteristics suggest that these molecules are also tracing jet-driven bow shocks \citep{2001Lee_jet_wind_model}. 
}
\end{figure*}

To have a closer look at the kinematics, we \REVII{made} the position-velocity (PV) diagrams \REVII{(Figure \ref{fig:PVtranss})} cut along the outflow axis with an average width of three pixels with a pixel size of 0\farcs{06} using CARTA 4.0 \citep[][]{2021Comrie_CARTA}. 
First, the gap of the CO (grey contours) at $\Delta v=0$~\kmpers\ indicates the missing flux due to interferometry at low velocities. 
\REVII{
Second, as shown in Figure \ref{fig:PVtranss} (k), the rapid growth of velocity $\Delta v$ with position offset $\Delta p$ from  $\Delta p=0\arcs$ to $0\farcs{3}$ and $-0\farcs{2}$ is significant in SiO. 
Finally, as shown in Figure \ref{fig:PVtranss} (a) to (g), the broadening of the velocity range localized at $\Delta p=0\farcs{3}$ and $-0\farcs{2}$ is significant in CH$_3$OH, H$_2$CO, OCS, and H$_2$S and tentative in $^{13}$CS. 
}
These characteristics suggest that these molecules (i.e., SiO, CH$_3$OH, H$_2$CO, OCS, H$_2$S, and $^{13}$CS) are more likely tracing jet-driven outflows rather than wide-angle winds \citep{2001Lee_jet_wind_model}. 

For the molecular transitions exhibiting two localized blobs, the northern ($\Delta p>0\arcs$) blobs have a larger velocity width than what the southern ($\Delta p<0\arcs$) blobs. 
As shown by the navy dashed polylines, most of the molecules including CH$_3$OH, OCS, and H$_2$S have comparable position offsets, indicating their possible common origin. 
The only exception is H$_2$CO, which has slightly higher position offset of $\sim$0\farcs{40}. 
Finally, for all transitions, the position offsets of the northern blobs ($\Delta p\sim0\farcs{3}$) are generally higher than the southern blobs ($-0\farcs{2}$), as also illustrated in Figure~\ref{fig:chn_CH3OH_Cycle3}. 

\begin{figure}[htb!]
\centering
\includegraphics[width=.99\linewidth]{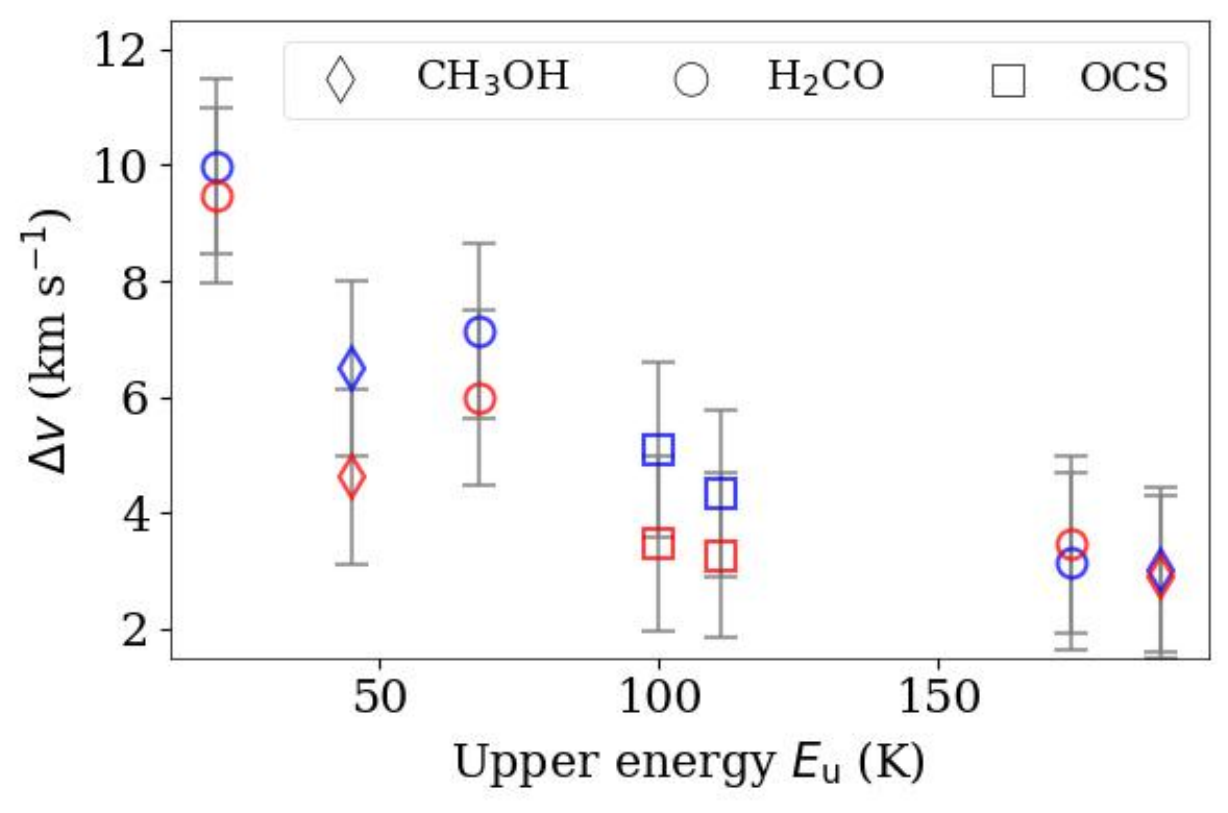}
\caption{\label{fig:trans_Eu} 
The velocity width versus upper energy for the molecular transitions in the two chemically rich regions. 
Only molecular species having multiple transitions are included. 
The blue and red markers represent the measurement at the blue-shifted (northern) and red-shifted (southern) lobes, respectively. 
The error bars are referred from the spectral channel widths ($\sim$~1.4~\kmpers.)
The velocity width seem to be anti-correlated with upper energies of the molecular transitions, possibly due to the different populations of each transition state. 
}
\end{figure}

Figure~\ref{fig:trans_Eu} shows the velocity widths of the blobs, which is defined as the maximum speed for each lobe in the PV diagrams (Figure~\ref{fig:PVtranss}) filtered out by a threshold of 5$\sigma$. 
The velocity widths seem to be anti-correlated with upper energies of the molecular transitions observed. 
As illustrated in Figure~\ref{fig:trans_Eu}, the velocity width of CH$_3$OH decreases from 5.6~\kmpers\ at \Eu~$=45$~K to 2.8\kmpers\ at \Eu~$=190$~K. 
For H$_2$CO, from the lowest (\Eu~$=21$~K) to the highest upper energy (\Eu~$=174$~K) transitions, the velocity width decreases from 9.8 to 2.8~\kmpers. 
\REVIII{These negative correlations could simply result from the different line intensities of the transitions having different upper energies, or it could possibly be due to the warmer gas at the inner regions. 
We favor the former based on the flat ratios between high and low transitions at high-velocity wings in Figure~\ref{fig:PVRatios}. 
In another word, the flat ratios imply that there is no significant change in the gas temperature.
}
However, the weak line intensities at high velocity, particularly for the high-upper-energy transitions, prevent us \REVII{from} justifying this.

\section{Discussion}
\label{sec:Disc}

\subsection{The Localized Chemically rich Features in HOPS 87}
\label{sec:disc:inventory}

Our findings suggest that the two chemically rich regions localized near the bases of the molecular jet are due to jet-driven outflows. 
The hosting molecular inventories in the shocked regions comprise H$_2$CO, CH$_3$OH, OCS, $^{13}$CS, and H$_2$S.

Finally, for all transitions, the position offsets of the northern blobs ($\Delta p\sim0\farcs{3}$) are generally higher than the southern blobs ($-0\farcs{2}$), as also illustrated in Figure~\ref{fig:chn_CH3OH_Cycle3}. 
\REVI{
Overall, most of the molecules have similar position offsets with each other for the northern lobe and the southern lobe individually. 
The only exception is H$_2$CO, which peaks at a slightly greater position offset.
Meanwhile, the northern (blue-shifted) lobe peaks at a larger position offset than the south (red-shifted) lobe for both H$_2$CO and all other transitions.
}
\subsubsection{The Origin of the Localized Chemically Rich Feature}

Two potential mechanisms are proposed for the bringing COMs from the ice phase to the gas phase in YSOs: thermal desorption and grain sputtering. 
For the former, ice sublimates at approximately 100~K and the frozen molecules are exclusively released into the gas phase \citep[e.g. ][]{2022Busch_themal_desorption}. 
In such case, the rotational temperature of the desorbed molecules is expected to be higher or comparable to the ice sublimation temperature at around 100 K. 
Even below ice sublimation temperature at $\sim$20--30 K, ``partial'' thermal desorption may possibly occur, specifically at the outermost layer of the icy mantle. 
This may result from lower binding energies (and consequently desorption energies), but the uncertain nature of these binding energies makes confirmation difficult \citep{2022Busch_themal_desorption}. 
As discussed by \citet{2022Busch_themal_desorption}, partial thermal desorption is inefficient compared to the complete thermal desorption (i.e., co-evaporation with ice). 

For the latter, the sputtering of grains requires shocks, commonly occurring when ejected/outflowing materials collide with the dense surrounding medium at high speeds. 
The collisions can \REVII{release H$_2$O as well as COMs from the ice phase to the gas phase} \citep[e.g., ][]{1997Caselli_grain-grain,2008Jimenez-Serra_grain_sputter}. 
The velocity dispersion of the molecular \REVII{lines} is expected to be broad due to the nature of shocks. 
Shocks can also heat up the materials, leading to the thermal desorption of molecules \citep[e.g., ][]{2017Miura_ShockChemistry}. 
After the shock front, the shocked materials begin to cool and are not necessarily warm \citep[e.g., ][]{2020James_L1157_ShockChem}.

Among the above two mechanisms (i.e., ice sublimation and grain sputtering) able to lead the richness of molecules, in particular those form within ice mantle such as CH$_3$OH and OCS, our findings favor the shock-induced grain sputtering picture \REVII{due to}
(1) the velocity spur structures (i.e., wide velocity dispersion) of the molecular tracers 
and
(2) the \REVII{coexistence} of SiO gas, an indicator of dust destruction \cite[e.g., ][]{2009Guillet_shock_SiO}. 
Although we do not see the \REVII{knots} of the typical tracers, CO and SiO, at the same position offset, we do see SiO knots at slightly higher position offsets ($\Delta p=\pm0\farcs{5}$), as shown in Figure \ref{fig:PVtranss} (k). 
We note that \citet{2024LeGouellec_Class0_2um} has detected molecular hydrogen lines in the 2~\micron\ band in HOPS 87, a spectral feature typically associated with shocks in protostellar outflow cavities.

\subsubsection{Comparisons with the counterpart hot corino}

\begin{figure*}[htb!]
\centering
\includegraphics[width=.85\textwidth]{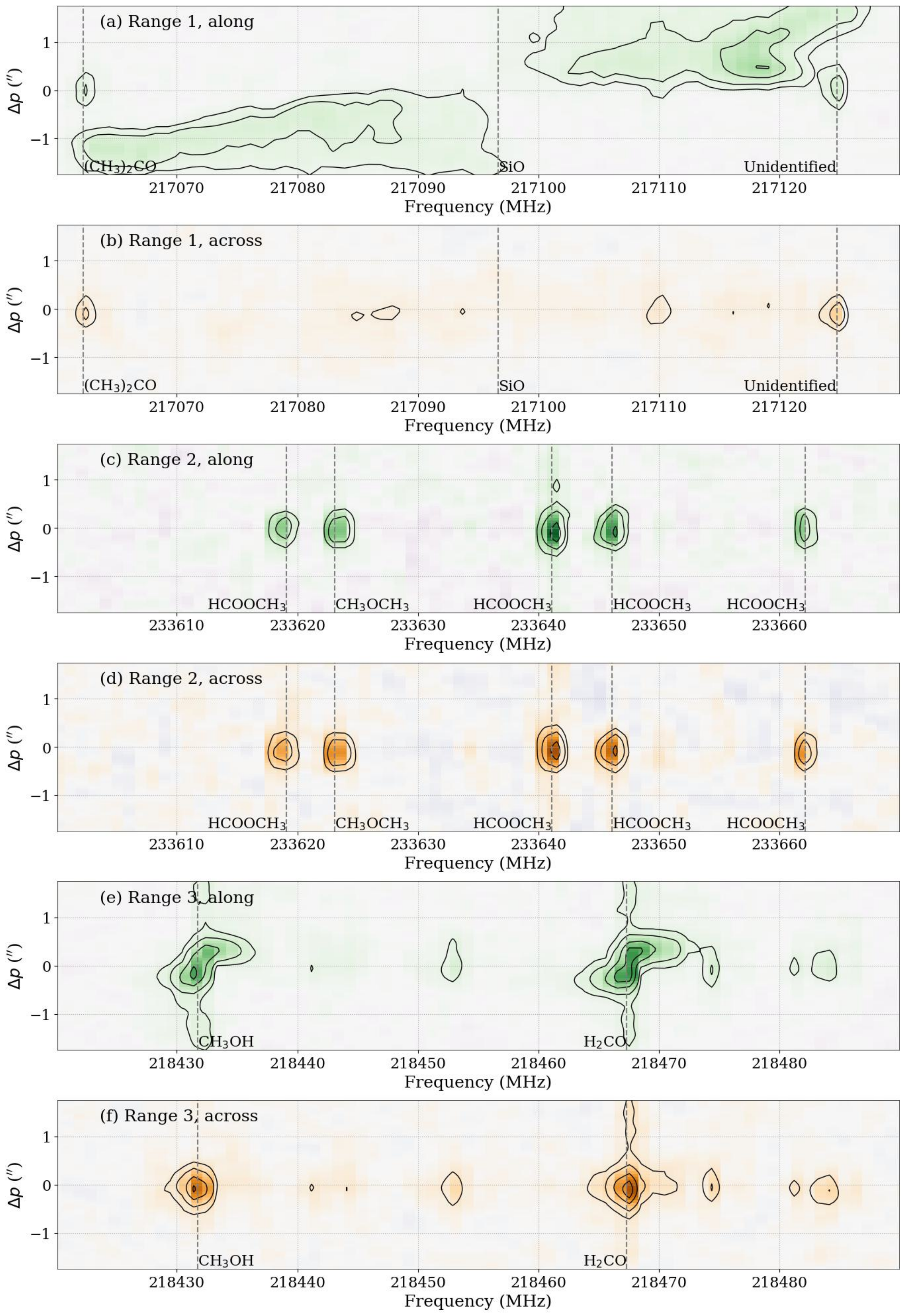}
\caption{\label{fig:PVcube}
PV diagrams \REVII{cut} along (panels a, c, and e) and across (panels b, d, and f) to the outflow axis for three selected frequency ranges. 
The x-axes are in terms of frequency. 
\REVII{The origin of the y-axes is the continuum peak.} 
The contours \REVII{represent} levels of [10, 20, 30, 40]$\sigma$. 
In panels (a) and (b), the typical outflowing gas tracer, SiO, is visible only in the cut along the outflow axis with a spur structure. 
Panels (c) and (d) show that COMs like HCOOCH$_3$ and CH$_3$OCH$_3$ display a single peak in both diagrams. 
In panels (e) and (f), CH$_3$OH and H$_2$CO present a single peak in the cut across the outflow axis, accompanied by a two-lobe feature in the cut along the outflow axis. 
}
\end{figure*}

To compare the chemical \REVII{compositions} of the hot corino and chemically rich shocked regions near the outflow origins, we extracted the PV diagrams with cuts along and across the outflow axis for molecules tracing \REVII{between} jet (SiO), hot corino zone (HCOOCH$_3$ and CH$_3$OCH$_3$), and both \REVII{regions} (CH$_3$OH and H$_2$CO) in Figure~\ref{fig:PVcube}. 
Note that the x-axes are in terms of frequency to cover multiple molecular transitions within each panel. 
As shown in \ref{fig:PVtranss} (a) and  (b), the typical outflowing gas tracer SiO only appears in the cut along the outflow axis and shows a spur structure. 
In \ref{fig:PVtranss} (c) and (d), COMs HCOOCH$_3$ and CH$_3$OCH$_3$ exhibit a single peak in both diagrams, suggesting that they are present only in the hot corino zone. 
In contrast, as illustrated in \ref{fig:PVtranss} (e) and (f), CH$_3$OH and H$_2$CO show a single peak on the cut across the outflow axis and a spur signature on the cut along the outflow axis. 
This indicates that some molecules (i.e., CH$_3$OH and H$_2$CO) residing in the hot corino zone are also present in the shocked regions.

\subsubsection{Implications}

On the basis that COMs are \REVII{thought to be} primarily formed within the icy mantle on dust grains and then released into the gas phase \citep[e.g., ][]{2009Herbst_COM_review,2016Yamamoto_book}, the observed gaseous COMs can \REVII{reveal a glimpse of} the ice composition. 
Therefore, the difference in gaseous COM composition between the hot corino zone and the jet-localized shocked regions in HOPS 87 suggests distinct ice compositions of the dust grains in these areas.
In the hot corino zone of HOPS 87, the dust grains have hosted a variety of COMs such as CH$_3$CHO, HCOOCH$_3$, and CH$_3$OCH$_3$. 
The presence of these larger COMs (compared to CH$_3$OH) indicates that the ice mantles have had a prolonged period for complex chemistry to occur.
In contrast, in the jet-localized shocked regions, the only firmly detected COM is CH$_3$OH. 
As the simplest COM and a precursor to larger COMs, the formation history of CH$_3$OH is relatively straightforward compared to larger COMs such as CH$_3$CHO. 
The formation of the former can be achieved via a series of hydrogenation with the frozen CO, and the formation of the latter requires more atoms/molecules as reactants \citep[e.g., ][]{2009Herbst_COM_review,2016Yamamoto_book}. 
This suggests that either the icy mantle, at least its top layer, on dust grains in the shocked region have not undergone as extensive chemical processing, or they have experienced cycles of ice sublimation and reformation, limiting the formation of larger COMs. 

Alternatively, the chemical composition could be affected by shocks. 
First, molecules could be destroyed by shocks \citep[e.g., ][]{2017Holdship_UCLCHEM_ShockChem}. 
For example, as speculated by \citet{2008Arce_L1157-B1_COMs}, the low CH$_3$CN/CH$_3$OH ratio in L1157 B1 shock may be due to the rapid destruction of CH$_3$CN. 
Similarly, in HOPS 87, the larger COMs in the jet-localized regions could possibly be destroyed. 
Second, shocks may enhance chemical reactions not commonly present in cold gas through the so-called ``shock chemistry," due to factors such as high temperature and sputtered molecular species \citep[e.g., ][]{1997Caselli_grain-grain,2019Burkhardt_ShockChemistry}. 
As proposed by \citet{2015Codella_L1157-B1_COMs}, a significant fraction of CH$_3$CHO \REVII{in L1157 B1 could form in the gas phase due to the shocks}. 
Meanwhile, the chemical composition could be sensitive to the evolution of shocks and further serve as a probe of the shock stage \citep[e.g., ][]{2016Burkhardt_L1157, 2019Burkhardt_ShockChemistry}. 
A more comprehensive investigation of the physical parameters (e.g., temperature) would help constrain the shock models. 
This would further improve our understanding of the dynamic interplay between the protostellar bipolar outflow, disk, inner core environment, and the surrounding medium \citep[e.g., ][]{1996Shu_Xwind_chondrites,2023Shang_unified_kinematics}.

\subsubsection{Comparison with L1157-B1}
\label{sec:disc:blob:L1157-B1}

L1157-B1 is a jet-driven bow shock recognized for housing an inventory of COMs \citep[e.g., ][]{2008Arce_L1157-B1_COMs,2014Mendoza_L1157-B1_COMs,2015Codella_L1157-B1_COMs,2017Lefloch_L1157-B1_COMs,2020Feng_L1157-B1_COMs,2020Spezzano_L1157-B1_COMs,2021Benedettini_L1157-B1_COMs}. 
\citet{2008Arce_L1157-B1_COMs} estimated the timescale of the outflows to be roughly less than 2000 years, which is too short for \REVII{their detected COMs to have formed primarily through gas-phase reactions}. 
Additionally, the shock is far from the associated central protostar (spatial offset~$=68\farcs{3}$, distance~$=250$ pc), rendering thermal desorption due to the central protostar unlikely. 
The authors concluded that the COMs more likely formed on the grain surface and then were expelled from the icy mantles by the shock.

For HOPS 87, we expect that the temperature of the molecules residing in the two jet-localized regions are warm based on the detection of transition having high upper energies (e.g., OCS 100 and 111 K transitions. )
In contrsast, the rotational temperatures of molecules within the L1157-B1 shock typically range from 20 to 50 K. 
For instance, the kinematic temperature of OCS was reported to be $46.8\pm3.4$~K by \citet{2019Holdship_L1527B1_S-bearing}. 
As demonstrated by \citet{2017Busquet_L1157-B1_DCN}, this relatively low temperature may possibly be a result of cooling after the shocks. 
On the basis of their demonstration, the relatively warm shocked region in HOPS 87 suggests that the cooling process is insignificant, consistent with the reported short dynamic age of the outflow.

\subsection{Suspected Component at the Bow Shocks}
In Figure~\ref{fig:PVtranss}, the CH$_3$OH (\Eu~$=45$ K) and H$_2$CO (\Eu~$=21$ K) transitions exhibit tentative components at $\Delta p\sim\pm$1\farcs{5}. 
Given their proximity to the edges of the bipolar outflow, we suspect that these emission components may trace the interaction surface between the bipolar outflow and the ambient quiescent material. 
This could be similar with the arc-like HCOOH emission pattern in Orion-KL observed by  \citet{2002Liu_OrionKL_HCOOC}. 

These components are expected to have low gas temperatures, given their distance (and \REVII{fewer} received photons consequently) from the central protostar and the non-detection in our data of high-upper-energy transitions such as OCS 100 and 111 K (Figure~\ref{fig:PVtranss}). 
Between the two potential associations, namely grain sputtering or thermal desorption, we favor the former due to the (expected) low temperature.
The absence of high-velocity wings, another signature of associations with shocks, may be attributed to the low SNR.
In this picture, these components would possibly exhibit chemical similarities to the chemically rich shocked regions such as L1157-B1 \citep[e.g., ][]{2008Arce_L1157-B1_COMs}, also associated with bow shocks.
A more sensitive investigation is necessary to reveal the nature of these components. 

\section{Conclusions}
\label{sec:Conclusions}

Using the ALMA Cycle 6 data from the ALMASOP project and archival ALMA Cycle 3 data, our findings include:

\begin{enumerate}  
  \item In the protostellar core HOPS 87, we discovered two chemically rich and localized regions containing CH$_3$OH, H$_2$CO, OCS, $^{13}$CS, and H$_2$S near the bases of the molecular bipolar outflow. 
  The position offsets of these regions are consistent with those of the bipolar outflow as traced by CO and SiO, with the northern (blue-shifted) lobe peaking at a larger position offset $\Delta p$ than the southern (red-shifted) lobe. 
  
  \item These localized chemically rich features are more likely due to shocks driven by jets rather than winds, based on the shapes of the PV diagrams. 
  Under the jet-driven shock scenario, the chemically rich and jet-localized regions are likely the knots often detected in jets. 
  
  \item Compared to the counterpart hot corino in HOPS 87, the composition of COMs in the shocked regions appears simpler. 
  We speculate that this simple composition is due to resident dust grains undergoing cycles of ice sublimation and subsequent reformation during their journey from the circumstellar envelope to the jet-localized region.
  Alternatively, it could result from the shock chemistry, and the chemical composition may further constrain the shock models. 

\end{enumerate}


\acknowledgments
We thank the two anonymous reviewers for valuable comments.
This paper makes use of the following ALMA data: ADS/JAO.ALMA\#2015.1.00341.S and ADS/JAO.ALMA\#2018.1.00302.S. ALMA is a partnership of ESO (representing its member states), NSF (USA) and NINS (Japan), together with NRC (Canada), NSTC and ASIAA (Taiwan), and KASI (Republic of Korea), in cooperation with the Republic of Chile. The Joint ALMA Observatory is operated by ESO, AUI/NRAO and NAOJ.
This work made use of Astropy:\footnote{http://www.astropy.org} a community-developed core Python package and an ecosystem of tools and resources for astronomy \citep{astropy:2013, astropy:2018, astropy:2022}. 
S.-Y. H. and C.-F.L. acknowledge grants from the National Science and Technology Council of Taiwan (110-2112-M-001-021-MY3 and 112-2112-M-001-039-MY3) and the Academia Sinica (Investigator Award AS-IA-108-M01).
LB gratefully acknowledges support by the ANID BASALproject FB210003. 
DJ is supported by NRC Canada and by an NSERC Discovery Grant.
Y.-L.Y. acknowledges support from Grant-in-Aid from the Ministry of Education, Culture, Sports, Science, and Technology of Japan (20H05845, 20H05844), and a pioneering project in RIKEN (Evolution of Matter in the Universe).

\software{
Astropy \citep{astropy:2013, astropy:2018, astropy:2022},
\texttt{CASA} \citep{casa:2022},
\texttt{CARTA}  \citep{2021Comrie_CARTA}, 
}

\clearpage
\appendix
\section{Molecular Transitions\label{appx:molec}}
\resetapptablenumbers

\begin{deluxetable*}{lrrll}[tbp!]
\label{tab:molec_info}
\caption{Transitions in this study}
\tablehead{
\colhead{Species} & \colhead{\Eu} & \colhead{\frest} & \colhead{Quantum Number} & \colhead{Database} \\
\colhead{} & \colhead{(K)} & \colhead{(MHz)} 
}
\startdata
CO & 17 & 230538 & v=0; J=2-1 & CDMS \\
SiO & 31 & 217105 & v=0; J=5-4 & CDMS \\
C$^{18}$O & 16 & 219560 & v=0; J=2-1 & CDMS \\
H$_2$CO & 21 & 218222 & J=3-2; Ka=0; Kc=3-2 & CDMS \\
H$_2$CO & 68 & 218476 & J=3-2; Ka=2; Kc=2-1 & CDMS \\
H$_2$CO & 68 & 218760 & J=3-2; Ka=2; Kc=1-0 & CDMS \\
H$_2$CO & 174 & 216569 & J=9; Ka=1; Kc=8-9 & CDMS \\
CH$_3$OH & 45 & 218440 & rovibSym=E; v=0; J=4-3; Ka=2-1; Kc=3-2 & CDMS \\
CH$_3$OH & 165 & 232419 & rovibSym=A1; v=0; J=10-9; Ka=2-3; Kc=8-7 & CDMS \\
CH$_3$OH & 190 & 232945 & rovibSym=E; v=0; J=10-11; Ka=3-2; Kc=7-9 & CDMS \\
CH$_3$CHO & 65 & 216582 & rovibSym=E; v=0; J=11-10; Ka=1; Kc=10-9 & JPL \\
HCOOCH$_3$ & 109 & 216110 & rovibSym=E; v=0; J=19-18; Ka=2; Kc=18-17 & JPL \\
H$_2$S & 84 & 216710 & v1=0; v2=0; v3=0; J=2; Ka=2-1; Kc=0-1 & CDMS \\
OCS & 100 & 218903 & J=18-17; v1=0; v2=0; v3=0 & CDMS \\
OCS & 111 & 231061 & J=19-18; v1=0; v2=0; v3=0 & CDMS \\
$^{13}$CS & 33 & 231221 & v=0; J=5-4 & CDMS \\
\enddata
\tablecomments{
\Eu\ is the energy of the upper state, \frest\ is the rest frequency. 
}
\end{deluxetable*}

\begin{figure*}[htb!]
\centering
\includegraphics[width=0.9\textwidth]{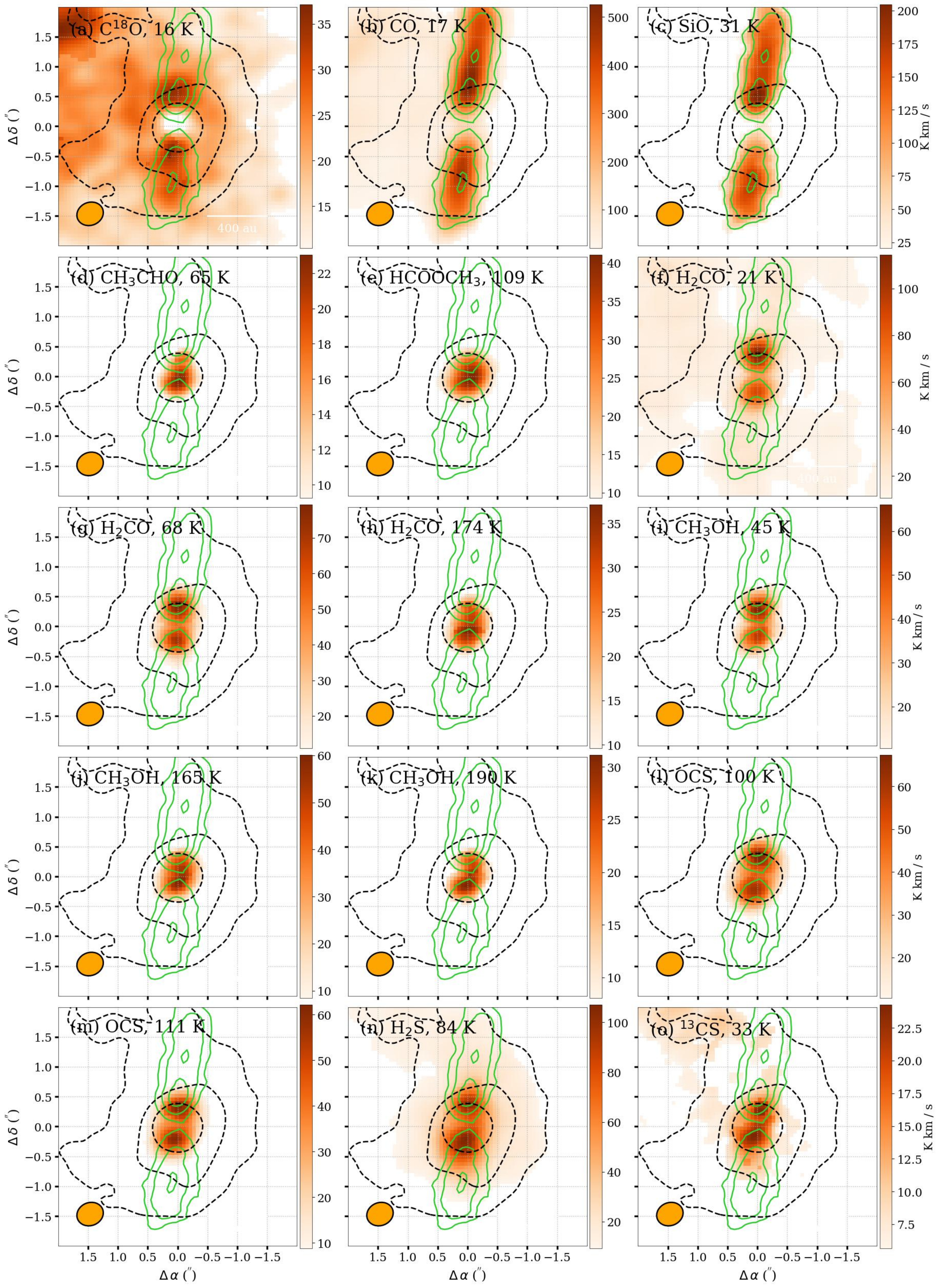}
\caption{\label{fig:mom0_all} 
Integrated intensity images of \REVII{all} transitions listed in Table~\ref{tab:molec_info}. 
The other captions follow Figure~\ref{fig:mom0_outflow}. 
\REVII{Note that some panels are repeated from Figure 1 but may be labeled differently}. 
\REVII{The} black and the green contours depict the 1.3~mm continuum and the SiO integrated intensity map, respectively. 
}
\end{figure*}

Table~\ref{tab:molec_info} shows the molecular transitions used in this study. 
In addition, Figure~\ref{fig:mom0_all} shows the integrated intensity maps of the transitions.

\clearpage
\section{Ratios between PV Diagram\label{appx:PV}}
\resetapptablenumbers

\begin{figure*}[htb!]
\centering
\includegraphics[width=.95\textwidth]{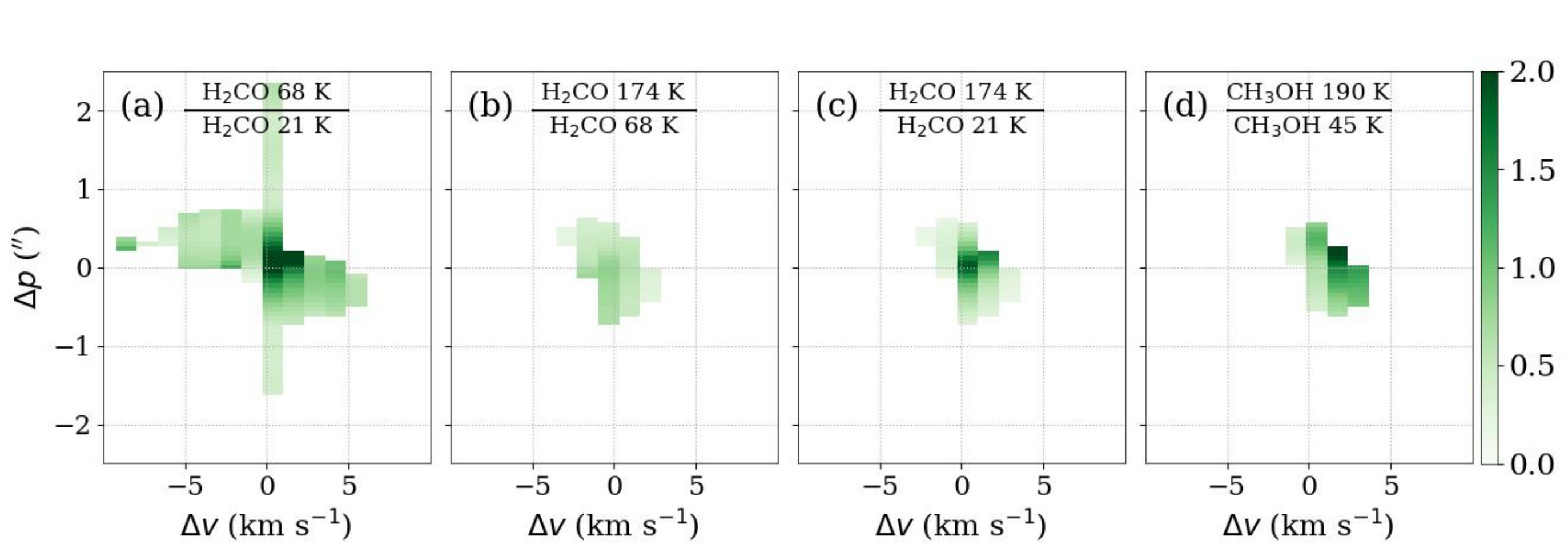}
\caption{\label{fig:PVRatios}
The PV diagrams of ratios between flux densities. 
The label in each panel denotes the molecular transitions for the numerator and denominator. 
The pixels having flux density smaller than 5$\sigma$ are exculded. 
}
\end{figure*}

Figure~\ref{fig:PVRatios} shows the PV diagrams of ratios between flux densities for selected transition pairs.

\bibliography{REFERENCE.bib}{}
\bibliographystyle{aasjournal}




\end{CJK*}
\end{document}